\documentclass[10pt,reqno]{amsart}
     \makeatletter
     \def\section{\@startsection{section}{1}%
     \z@{.7\linespacing\@plus\linespacing}{.5\linespacing}%
     {\bfseries
     \centering
     }}
     \def\@secnumfont{\bfseries}
     \makeatother
\usepackage{amsmath}
\usepackage{amssymb}
\usepackage[latin1]{inputenc}

\newcommand*{\cA}{\mathcal{A}}
\newcommand*{\cB}{\mathcal{B}}
\newcommand*{\cC}{\mathcal{C}}
\newcommand*{\C}{\mathbb{C}}
\newcommand*{\Z}{\mathbb{Z}}
\newcommand*{\N}{\mathbb{N}}
\newcommand*{\1}{\mathbf{1}}
\newcommand*{\R}{\mathbb{R}}

\newcommand{\abs}[1]{\left|#1\right|}
\newcommand{\norm}[1]{\left\|#1\right\|}

\renewcommand{\and}{\text{ and }}

\newcommand{\maps}{\rightarrow}

\newcommand{\bmx}{\mathbf{x}}

\newcommand{\bmid}{\mathbf{Id}}
\newcommand{\bmk}{\mathbf{K}}
\newcommand{\bml}{\mathbf{L}}
\newcommand{\bmn}{\mathbf{N}}
\newcommand{\bmni}{\mathbf{N}^{-1}}
\newcommand{\bmy}{\mathbf{y}}

\newcommand{\bmf}{\mathbf{f}}
\newcommand{\bmg}{\mathbf{g}}

\newcommand{\Nexp}{\text{Nexp}}

\newtheorem{theorem}{Theorem}[section]
\newtheorem{lemma}[theorem]{Lemma}
\newtheorem{proposition}[theorem]{Proposition}
\newtheorem{corollary}[theorem]{Corollary}
\theoremstyle{definition}
\newtheorem{definition}[theorem]{Definition}
\newtheorem{example}[theorem]{Example}
\theoremstyle{remark}
\newtheorem{remark}[theorem]{Remark}
\numberwithin{equation}{section}
\begin{document}

\title[Charged Particle as Hida Distribution]{The Feynman integrand for the Charged Particle in a Constant Magnetic field as White Noise Distribution}

\author[Wolfgang Bock]{Wolfgang Bock}
\address{Functional Analysis and Stochastic Analysis Group, \\
Department of Mathematics, \\
University of Kaiserslautern, 67653 Kaiserslautern, Germany}
\email{bock@mathematik.uni-kl.de}
\urladdr{http://www.mathematik.uni-kl.de/$\sim$bock}

\author[Martin Grothaus]{Martin Grothaus }
\address{Functional Analysis and Stochastic Analysis Group, \\
Department of Mathematics, \\
University of Kaiserslautern, 67653 Kaiserslautern, Germany}
\email{bock@mathematik.uni-kl.de}
\urladdr{http://www.mathematik.uni-kl.de/$\sim$grothaus}

\author[Sebastian Jung]{Sebastian Jung}
\address{Functional Analysis and Stochastic Analysis Group, \\
Department of Mathematics, \\
University of Kaiserslautern, 67653 Kaiserslautern, Germany}
\email{sjung@mathematik.uni-kl.de }
\urladdr{http://www.mathematik.uni-kl.de/$\sim$wwwfktn}



\begin{abstract}
The concepts of Feynman integrals in white noise analysis are used to realize the Feynman integrand for a charged particle in a constant magnetic field as a Hida distribution. For this purpose we identify the velocity dependent potential as a so called generalized Gauss kernel. 
\end{abstract}

\maketitle

\section{Introduction}
As an alternative approach to quantum mechanics Feynman introduced the concept of path integrals (\cite{F48,Fe51,FeHi65}), which was developed into an extremely useful tool in many branches of theoretical physics. In this article we use concepts for realizing Feynman integrals in the framework of white noise analysis. The Feynman integral for a particle moving from $0$ at time $0$ to $\mathbf{y} \in \R^d$ at time $t$ under the potential $V$ is given by 
\begin{equation}\label{eqnfey} 
{\rm N} \int_{\mathbf{x}(0)=0, \mathbf{x}(t)=\mathbf{y}} \int \exp\left(\frac{i}{\hbar} \int_0^t \frac{1}{2}m\dot{\mathbf{x}}^2 -V(\mathbf{x},\dot{\mathbf{x}}) \, d\tau \right) \prod_{0<\tau<t} d\mathbf{x}(\tau),\quad \hbar = \frac{h}{2\pi}.
\end{equation}
Here $h$ is Planck's constant, and the integral is thought of being over all paths with $\mathbf{x}(0)=0$ and $\mathbf{x}(t)=\mathbf{y}$.\\
In the last fifty years there have been many approaches for giving a mathematically rigorous meaning to the Feynman integral by using e.g.~ analytic continuation,limits of finite dimensional approximations or Fresnel integrals. Instead of giving a complete list of publications concerning Feynman integrals we refer to \cite{AHKM08} and the references therein.  Here we choose a white noise approach.
white noise analysis is a mathematical framework which offers generalizations of concepts from finite-dimensional analysis, like differential operators and Fourier transform to an infinite-dimensional setting. We give a brief introduction to white noise analysis in Section 2, for more details see \cite{Hid80,HKPS93,Ob94,BK95,Kuo96}. Of special importance in white noise analysis are spaces of generalized functions and their characterizations. In this article we choose the space of Hida distributions, see Section 2.\\
The idea of realizing Feynman integrals within the white noise framework goes back to \cite{HS83}. There the authors used exponentials of quadratic (generalized) functions in order to give meaning to the Feynman integral in configuration space representation
\begin{equation*}
{\rm N}\int_{\mathbf{x}(0) =0, \mathbf{x}(t)=y} \exp\left(\frac{i}{\hbar} S(\mathbf{x}) \right) \, \prod_{0<\tau<t} \, d\mathbf{x}(\tau) ,\quad \hbar = \frac{h}{2\pi},
\end{equation*}
with the classical action $S(\mathbf{x})= \int_0^t \frac{1}{2} m \dot{\mathbf{x}}^2 -V(\mathbf{x})\, d\tau$.
We use these concepts of quadratic actions in white noise analysis, which were further developed in \cite{GS98a} and \cite{BG10} to give a rigorous meaning to the Feynman integrand 
\begin{multline}\label{integrandpot}
I_V = {\rm Nexp}\left( \frac{i}{\hbar}\int_0^t  \frac{m}{2} \dot{\mathbf{x}}(\tau)^2  d\tau +\frac{1}{2}\int_0^t \dot{\mathbf{x}}(\tau)^2 d\tau\right)\\
\times \exp\left(-\frac{i}{\hbar} \int_0^t V(\mathbf{x}(\tau),\dot{\mathbf{x}}(\tau),\tau) \, d\tau\right) \cdot \delta_0(\mathbf{x}(t)-y)
\end{multline}
as a Hida distribution. In this expression the sum of the first and the third integral in the exponential is the action $S(\mathbf{x},\mathbf{\dot{x}})$, and the delta function (Donsker's delta) serves to pin trajectories to $\mathbf{y}$ at time $t$. The second integral is introduced to simulate the locally Lebesgue integral by a local compensation of the fall-off of the Gaussian reference measure $\mu$. 
Furthermore we use a two-dimensional Brownian motion starting in $0$ as the path i.e.
\begin{equation}\label{varchoice}
\mathbf{x}(\tau)=\sqrt{\frac{\hbar}{m}}\mathbf{B}(\tau).
\end{equation}
The construction is done in terms of the $T$-transform (infinite-dimensional version of the Fourier transform w.r.t~a Gaussian measure), which characterizes Hida distributions, see Theorem \ref{charthm}. At the same time, the $T$-transform of the constructed Feynman integrands provides us with their generating functional. Finally using the generating functional, we can show that the generalized expectation (generating functional at zero) gives the Greens function to the corresponding Schrödinger equation. \\
In this article we consider the potential given by the action of a constant magnetic field to a moving particle. From classical physics it is well-known, that a magnetic field is influencing the so-called Lorentz force on a charged particle moving through this field. The corresponding potential term of a charged particle moving in the $(1,2)$-plane is given by
$$(\bmx,\dot{\bmx}) \mapsto V_{\rm{mag}}(\bmx,\dot{\bmx})= -\frac{q H_3}{c} \left(x_1\dot{x_2}-\dot{x_1}x_2\right),$$
where $q$ is the charge, $H_3$ the strength of the magnetic field vector orthogonal to the $(1,2)$-plane and $c$ the speed of light.

These are the core results of this article:
\begin{itemize}
\item The concepts of generalized Gauss kernels from \cite{GS98a} and \cite{BG10} are used to construct the Feynman integrand for a charged particle in a constant magnetic field as a Hida distribution, see Theorem \ref{magnetictheorem}.
\item The results in Theorem \ref{magnetictheorem} provide us with the generating functional for a charged particle in a constant magnetic field. 
\item The generalized expectations (generating functional at zero) yields the Greens functions to the corresponding Schrödinger equation.
\end{itemize}

\section{White Noise Analysis}
\subsection{Gel'fand Triples}
Starting point is the Gel'fand triple $S_d(\R) \subset L^2_d(\R,dx) \subset S'_d(\R)$ of the $\R^d$-valued, $d \in \N$, Schwartz test functions and tempered distributions with the Hilbert space of (equivalence classes of) $\R^d$-valued square integrable functions w.r.t.~the Lebesgue measure as central space (equipped with its canonical inner product $(\cdot, \cdot)$ and norm $\|\cdot\|$), see e.g.~ \cite[Exam.~11]{W95}.
Since $S_d(\R)$ is a nuclear space, represented as projective limit of a decreasing chain of Hilbert spaces $(H_p)_{p\in \N}$, see e.g.~\cite[Chap.~2]{RS75a} and \cite{GV68}, i.e.~
\begin{equation*}
S_d(\R) = \bigcap_{p \in \N} H_p,
\end{equation*}
we have that $S_d(\R)$ is a countably Hilbert space in the sense of Gel'fand and Vilenkin \cite{GV68}. We denote the inner product and the corresponding norm on $H_p$ by $(\cdot,\cdot)_p$ and $\|\cdot\|_p$, respectively, with the convention $H_0 = L^2_d(\R, dx)$.
Let $H_{-p}$ be the dual space of $H_p$ and let $\langle \cdot , \cdot \rangle$ denote the dual pairing on $H_{p} \times H_{-p}$. $H_{p}$ is continuously embedded into $L^2_d(\R,dx)$. By identifying $L_d^2(\R,dx)$ with its dual $L_d^2(\R,dx)'$, via the Riesz isomorphism, we obtain the chain $H_p \subset L_d^2(\R, dx) \subset H_{-p}$.
Note that $\displaystyle S'_d(\R)= \bigcup_{p\in \N} H_{-p}$, i.e.~$S'_d(\R)$ is the inductive limit of the increasing chain of Hilbert spaces $(H_{-p})_{p\in \N}$, see  e.g.~\cite{GV68}.
We denote the dual pairing of $S_d(\R)$ and $S'_d(\R)$ also by $\langle \cdot , \cdot \rangle$. Note that its restriction on $S_d(\R) \times L_d^2(\R, dx)$ is given by $(\cdot, \cdot )$.
We also use the complexifications of these spaces denoted with the subindex $\C$ (as well as their inner products and norms). The dual pairing we extend in a bilinear way. Hence we have the relation 
\begin{equation*}
\langle g,f \rangle = (\mathbf{g},\overline{\mathbf{f}}), \quad \mathbf{f},\mathbf{g} \in L_d^2(\R)_{\C},
\end{equation*}
where the overline denotes the complex conjugation.
\subsection{White Noise Spaces}
We consider on $S_d' (\R)$ the $\sigma$-algebra $\cC_{\sigma}(S_d' (\R))$ generated by the cylinder sets $\{ \omega \in S_d' (\R) | \langle \xi_1, \omega \rangle \in F_1, \dots ,\langle \xi_n, \omega \rangle \in F_n\} $, $\xi_i \in S_d(\R)$, $ F_i \in \cB(\R),\, 1\leq i \leq n,\, n\in \N$, where $\cB(\R)$ denotes the Borel $\sigma$-algebra on $\R$.\\
\noindent The canonical Gaussian measure $\mu$ on $C_{\sigma}(S_d'(\R))$ is given via its characteristic function
\begin{eqnarray*}
\int_{S_d' (\R)} \exp(i \langle {\bf f}, \boldsymbol{\omega} \rangle ) d\mu(\boldsymbol{\omega}) = \exp(- \tfrac{1}{2} \| {\bf f}\|^2 ), \;\;\; {\bf f} \in S_d(\R),
\end{eqnarray*}
\noindent by the theorem of Bochner and Minlos, see e.g.~\cite{Mi63}, \cite[Chap.~2 Theo.~1.~11]{BK95}. The space $(S_d'(\R),\cC_{\sigma}(S_d'(\R)), \mu)$ is the ba\-sic probability space in our setup.
The cen\-tral Gaussian spa\-ces in our frame\-work are the Hil\-bert spaces $(L^2):= L^2(S_d'(\R),$ $\cC_{\sigma}(S_d' (\R)),\mu)$ of complex-valued square in\-te\-grable func\-tions w.r.t.~the Gaussian measure $\mu$.\\
Within this formalism a representation of a d-dimensional Brownian motion is given by 
\begin{equation}\label{BrownianMotion}
{\bf B}_t ({\boldsymbol \omega}) :=(B_t(\omega_1), \dots, B_t(\omega_d)):= ( \langle  \1_{[0,t)},\omega_1 \rangle, \dots  \langle  \1_{[0,t)},\omega_d \rangle),\end{equation}
with ${\boldsymbol \omega}=(\omega_1,\dots, \omega_d) \in S'_d(\R),\quad t \geq 0,$
in the sense of an $(L^2)$-limit. Here $\1_A$ denotes the indicator function of a set $A$. 

\subsection{The Hida triple}

Let us now consider the Hilbert space $(L^2)$ and the corresponding Gel'fand triple
\begin{equation*}
(S) \subset (L^2) \subset (S)'.
\end{equation*}
Here $(S)$ denotes the space of Hida test functions and $(S)'$ the space of Hida distributions. In the following we denote the dual pairing between elements of $(S)$ and $(S)'$ by $\langle \! \langle \cdot , \cdot \rangle \!\rangle$. 
Instead of reproducing the construction of $(S)'$ here we give its characterization in terms of the $T$-transform.\\
\begin{definition}
We define the $T$-transform of $\Phi \in (S)'$ by
\begin{equation*}
T\Phi({\bf f}) := \langle\!\langle  \exp(i \langle {\bf f}, \cdot \rangle),\Phi \rangle\!\rangle, \quad  {\bf f}:= ({ f_1}, \dots ,{ f_d }) \in S_{d}(\R).
\end{equation*}
\end{definition}

\begin{remark}
\begin{itemize}
\item[(i)] Since $\exp(i \langle {\bf f},\cdot \rangle) \in (S)$ for all $f \in S_d(\R)$, the $T$-transform of a Hida distribution is well-defined.
\item[(ii)] For ${\bf f} = 0$ the above expression yields $\langle\!\langle \Phi, 1 \rangle\!\rangle$, therefore $T\Phi(0)$ is called the generalized expectation of $\Phi \in (S)'$.
\end{itemize}
\end{remark}

\noindent In order to characterize the space $(S)'$ by the $T$-transform we need the following definition.

\begin{definition}
A mapping $F:S_{d}(\R) \to \C$ is called a {\emph U-functional} if it satisfies the following conditions:
\begin{itemize}
\item[U1.] For all ${\bf{f, g}} \in S_{d}(\R)$ the mapping $\R \ni \lambda \mapsto F(\lambda {\bf f} +{\bf g} ) \in \C$ has an analytic continuation to $\lambda \in \C$ ({\bf{ray analyticity}}).
\item[U2.] There exist constants $0<C,D<\infty$ and a $p \in \N_0$ such that 
\begin{equation*}
|F(z{\bf f})|\leq C\exp(D|z|^2 \|{\bf f} \|_p^2), 
\end{equation*}
for all $z \in \C$ and ${\bf f} \in S_{d}(\R)$ ({\bf{growth condition}}).
\end{itemize}
\end{definition}

\noindent This is the basis of the following characterization theorem. For the proof we refer to \cite{PS91,Kon80,HKPS93,KLPSW96}.

\begin{theorem}\label{charthm}
A mapping $F:S_{d}(\R) \to \C$ is the $T$-transform of an element in $(S)'$ if and only if it is a U-functional.
\end{theorem}
Theorem \ref{charthm} enables us to discuss convergence of sequences of Hida distributions by considering the corresponding $T$-transforms, i.e.~ by considering convergence on the level of U-functionals. The following corollary is proved in \cite{PS91,HKPS93,KLPSW96}.

\begin{corollary}\label{seqcor}
Let $(\Phi_n)_{n\in \N}$ denote a sequence in $(S)'$ such that:
\begin{itemize}
\item[(i)] For all ${\bf f} \in S_{d}(\R)$, $((T\Phi_n)({\bf f}))_{n\in \N}$ is a Cauchy sequence in $\C$.
\item[(ii)] There exist constants $0<C,D<\infty$ such that for some $p \in \N_0$ one has 
\begin{equation*}
|(T\Phi_n)(z{\bf f })|\leq C\exp(D|z|^2\|{\bf f}\|_p^2)
\end{equation*}
for all ${\bf f} \in S_{d}(\R),\, z \in \C$, $n \in \N$.
\end{itemize}
Then $(\Phi_n)_{n\in \N}$ converges strongly in $(S)'$ to a unique Hida distribution.
\end{corollary}

\begin{example}[Vector valued white noise]
\noindent Let $\,{\bf{B}}(t)$, $t\geq 0$, be the $d$-di\-men\-sional Brow\-nian motion as in \eqref{BrownianMotion}. 
Consider $$\frac{{\bf{B}}(t+h,\boldsymbol{\omega}) - {\bf{B}}(t,\boldsymbol{\omega})}{h} = (\langle \frac{\1_{[t,t+h)}}{h} , \omega_1 \rangle , \dots (\langle \frac{\1_{[t,t+h)}}{h} , \omega_d \rangle),\quad h>0.$$ 
Then in the sense of Corollary \ref{seqcor} it exists
\begin{eqnarray*}
\langle {\boldsymbol\delta_t}, {\boldsymbol \omega} \rangle := (\langle \delta_t,\omega_1 \rangle, \dots ,\langle \delta_t,\omega_d \rangle):= \lim_{h\searrow 0} \frac{{\bf{B}}(t+h,\boldsymbol{\omega}) - {\bf{B}}(t,\boldsymbol{\omega})}{h}.
\end{eqnarray*}
Of course for the left derivative we get the same limit. Hence it is natural to call the generalized process $\langle {\boldsymbol\delta_t}, {\boldsymbol \omega} \rangle$, $t\geq0$ in $(S)'$ vector valued white noise. One also uses the notation ${\boldsymbol \omega}(t) =\langle{\boldsymbol\delta_t}, {\boldsymbol \omega} \rangle$, $t\geq 0$. 
\end{example}

Another useful corollary of Theorem \ref{charthm} concerns integration of a family of generalized functions, see \cite{PS91,HKPS93,KLPSW96}.

\begin{corollary}\label{intcor}
Let $(\Lambda, \cA, \nu)$ be a measure space and $\Lambda \ni\lambda \mapsto \Phi(\lambda) \in (S)'$ a mapping. We assume that its $T$--transform $T \Phi$ satisfies the following conditions:
\begin{enumerate}
\item[(i)] The mapping $\Lambda \ni \lambda \mapsto T(\Phi(\lambda))({\bf f})\in \C$ is measurable for all ${\bf f} \in S_d(\R)$.
\item[(ii)] There exists a $p \in \N_0$ and functions $D \in L^{\infty}(\Lambda, \nu)$ and $C \in L^1(\Lambda, \nu)$ such that 
\begin{equation*}
   \abs{T(\Phi(\lambda))(z{\bf f})} \leq C(\lambda)\exp(D(\lambda) \abs{z}^2 \norm{{\bf f}}^2), 
\end{equation*}
for a.e.~$ \lambda \in \Lambda$ and for all ${\bf f} \in S_d(\R)$, $z\in \C$.
\end{enumerate}
Then, in the sense of Bochner integration in $H_{-q} \subset (S)'$ for a suitable $q\in \N_0$, the integral of the family of Hida distributions is itself a Hida distribution, i.e.~$\!\displaystyle \int_{\Lambda} \Phi(\lambda) \, d\nu(\lambda) \in (S)'$ and the $T$--transform interchanges with integration, i.e.~
\begin{equation*}
   T\left( \int_{\Lambda} \Phi(\lambda) \, d\nu(\lambda) \right)(\mathbf{f}) =
   	\int_{\Lambda} T(\Phi(\lambda))(\mathbf{f}) \, d\nu(\lambda), \quad \mathbf{f} \in S_d(\R).
\end{equation*}
\end{corollary}

Based on the above theorem, we introduce the following Hida distribution.
\begin{definition}
\label{D:Donsker} 
We define Donsker's delta at $x \in \R$ corresponding to $0 \neq {\boldsymbol\eta} \in L_{d}^2(\R)$ by
\begin{equation*}
   \delta_0(\langle {\boldsymbol\eta},\cdot \rangle-x) := 
   	\frac{1}{2\pi} \int_{\R} \exp(i \lambda (\langle {\boldsymbol\eta},\cdot \rangle -x)) \, d \lambda
\end{equation*}
in the sense of Bochner integration, see e.g.~\cite{HKPS93,LLSW94,W95}. Its $T$--transform in ${\bf f} \in S_d(\R)$ is given by
\begin{equation*}
   T(\delta_0(\langle  {\boldsymbol\eta},\cdot \rangle-x)({\bf f}) 
   	= \frac{1}{\sqrt{2\pi \langle {\boldsymbol\eta}, {\boldsymbol\eta}\rangle}} \exp\left( -\frac{1}{2\langle {\boldsymbol\eta},{\boldsymbol\eta} \rangle}(i\langle {\boldsymbol\eta},{\bf f} \rangle - x)^2 -\frac{1}{2}\langle {\bf f},{\bf f}\rangle \right), \, \, \mathbf{f} \in S_d(\R).
\end{equation*}
\end{definition}

\subsection{Generalized Gauss Kernels}
Here we review a special class of Hida distributions which are defined by their T-transform, see e.g.~\cite{HS83},\cite{HKPS93},\cite{GS98a}. Proofs and more details for can be found in \cite{BG10}. Let $\mathcal{B}$ be the set of all continuous bilinear mappings $B:S_{d}(\R) \times S_{d}(\R) \to \C$. Then the functions
\begin{equation*}
S_d(\R)\ni \mathbf{f} \mapsto \exp\left(-\frac{1}{2} B({\bf f},{\bf f})\right) \in \C
\end{equation*}
for all $B\in \mathcal{B}$ are U-functionals. Therefore, by using the characterization of Hida distributions in Theorem \ref{charthm},
the inverse T-transform of these functions 
\begin{equation*}
\Phi_B:=T^{-1} \exp\left(-\frac{1}{2} B\right)
\end{equation*}
are elements of $(S)'$.

\begin{definition}\label{GGK}
The set of {\bf{generalized Gauss kernels}} is defined by
\begin{equation*}
GGK:= \{ \Phi_B,\; B\in \mathcal{B} \}.
\end{equation*}
\end{definition}

\begin{example}{\cite{GS98a}} \label{Grotex} We consider a symmetric trace class operator $\mathbf{K}$ on $L^2_{d}(\R)$ such that $-\frac{1}{2}<\mathbf{K}\leq 0$, then
\begin{align*}
\int_{S'_{d}(\R)} \exp\left(- \langle \omega,\mathbf{K} \omega\rangle \right) \, d\mu(\boldsymbol{\omega}) 
= \left( \det(\mathbf{Id +2K})\right)^{-\frac{1}{2}} < \infty.
\end{align*}
For the definition of $\langle \cdot,\mathbf{K} \cdot \rangle$ see the remark below.
Here $\mathbf{Id}$ denotes the identity operator on the Hilbert space $L^2_{d}(\R)$, and $\det(\mathbf{A})$ of a symmetric trace class operator $\mathbf{A}$ on $L^2_{d}(\R)$ denotes the infinite product of its eigenvalues, if it exists. In the present situation we have $\det(\mathbf{Id +2K})\neq 0$.
There\-fore we obtain that the exponential $g= \exp(-\frac{1}{2} \langle \cdot,\mathbf{K} \cdot \rangle)$ is square-integrable and its T-transform is given by 
\begin{equation*}
Tg({\bf f}) = \left( \det(\mathbf{Id+K}) \right)^{-\frac{1}{2}} \exp\left(-\frac{1}{2} ({\bf f}, \mathbf{(Id+K)^{-1}} {\bf f})\right), \quad {\bf f} \in S_{d}(\R).
\end{equation*}
Therefore $\left( \det(\mathbf{Id+K}) \right)^{\frac{1}{2}}g$ is a generalized Gauss kernel.
\end{example}

\begin{remark}
\begin{itemize}
\item[i)]\label{traceL2} Since a trace class operator is compact, see e.g.~\cite{RS75a}, we have that $\mathbf{K}$ in the above example is diagonalizable, i.e.~
\begin{equation*}
\mathbf{K}\mathbf{f} = \sum_{k=1}^{\infty} k_n (\mathbf{f},\mathbf{e}_n)\mathbf{e}_n, \quad \mathbf{f} \in L_d^2(\R,dx),
\end{equation*}
where $(\mathbf{e}_n)_{n\in \N}$ denotes an eigenbasis of the corresponding eigenvalues $(k_n)_{n\in \N}$ with $k_n \in (-\frac{1}{2}, 0 ]$, for all $n \in \N$. Since $K$ is compact, we have that $\lim\limits_{n\to \infty} k_n =0$ and since $\mathbf{K}$ is trace class we also have $\sum_{n=1}^{\infty} (\mathbf{e}_n, -\mathbf{K} \mathbf{e}_n)< \infty$. We define for ${\boldsymbol \omega }\in S_d'(\R)$
\begin{eqnarray*} 
- \langle {\boldsymbol \omega }, \mathbf{K} {\boldsymbol \omega } \rangle := \lim_{N \to \infty} \sum_{n=1}^N \langle \mathbf{e}_n, {\boldsymbol \omega }\rangle (-k_n)\langle \mathbf{e}_n,{\boldsymbol \omega } \rangle. 
\end{eqnarray*}
Then as a limit of measurable functions ${\boldsymbol \omega } \mapsto -\langle {\boldsymbol \omega }, \mathbf{K} {\boldsymbol \omega } \rangle$  is measurable and hence 
\begin{eqnarray*} 
\int\limits_{S_d'(\R)} \exp(-  \langle {\boldsymbol \omega }, \mathbf{K} {\boldsymbol \omega }\rangle ) \, d\mu({\boldsymbol \omega }) \in [0, \infty].
\end{eqnarray*}
The explicit formula for the $T$-transform and expectation then follow by a straightforward calculation with help of the above limit procedure. 
\item[ii)] In the following, if we apply operators or bilinear forms defined on $L^2_d(\R)$ to generalized functions from $S'_d(\R)$, we are always having in mind the interpretation as in \ref{traceL2}.
\end{itemize}
\end{remark}
\begin{definition}\label{D:Nexp}\cite{BG10}$\;$
Let $\mathbf{K}: L^2_d(\R, dx)_{\C} \to L^2_d(\R, dx)_{\C}$ be linear and continuous such that:
\begin{itemize}
\item[(i)] $\mathbf{Id+K}$ is injective. 
\item[(ii)] There exists $p \in \N_0$ such that $(\mathbf{Id+K})(L^2_{d}(\R,\,dx)_{\C}) \subset H_{p,\C}$ is dense.
\item[(iii)] There exist $q \in\N_0$ such that $\mathbf{(Id+K)^{-1}} :H_{p,\C} \to H_{-q,\C}$ is continuous with $p$ as in (ii).
\end{itemize}
Then we define the normalized exponential
\begin{equation}\label{Nexp}
{\rm{Nexp}}(- \frac{1}{2} \langle \cdot ,\mathbf{K} \cdot \rangle)
\end{equation}
by
\begin{equation*}
T({\rm{Nexp}}(- \frac{1}{2} \langle \cdot ,\mathbf{K} \cdot \rangle))({\bf f}) := \exp(-\frac{1}{2} \langle {\bf f}, \mathbf{(Id+K)^{-1}} {\bf f} \rangle),\quad {\bf f} \in S_d(\R).
\end{equation*}
\end{definition}

\begin{remark}
The "normalization" of the exponential in the above definition can be regarded as a division of a divergent factor. In an informal way one can write
\begin{multline*}
T({\rm{Nexp}}(- \frac{1}{2} \langle \cdot ,\mathbf{K} \cdot \rangle))({\mathbf f})=\frac{T(\exp(- \frac{1}{2} \langle \cdot ,\mathbf{K} \cdot \rangle))(\mathbf{f})}{T(\exp(- \frac{1}{2} \langle \cdot ,\mathbf{K} \cdot \rangle))(0)}\\
=\frac{T(\exp(- \frac{1}{2} \langle \cdot ,\mathbf{K} \cdot \rangle))(\mathbf{f})}{\sqrt{\det(\mathbf{Id+K})}} , \quad {\bf f} \in S_d(\R), 
\end{multline*}
i.e.~ if the determinant in the Example \ref{Grotex} above is not defined, we can still define the normalized exponential by the T-transform without the diverging prefactor. The assumptions in the above definition then guarantee the existence of the generalized Gauss kernel in \eqref{Nexp}.
\end{remark}

\begin{example}\label{pointprod}
	For sufficiently "nice" operators $\mathbf{K}$ and $\mathbf{L}$ on $L^2_{d}(\R)_{\C}$ we can define the product 
			\begin{equation*}
				{\rm{Nexp}}\big( - \frac{1}{2} \langle \cdot,\mathbf{K} \cdot \rangle  \big) \cdot \exp\big(-\frac{1}{2} \langle \cdot,\mathbf{L}\cdot \rangle \big)
			\end{equation*}
	of two square-integrable functions. Its $T$-transform is then given by 
			\begin{multline*}
				T\Big({\rm{Nexp}}( - \frac{1}{2} \langle \cdot,\mathbf{K} \cdot\rangle ) \cdot \exp( - \frac{1}{2} \langle \cdot,\mathbf{L} \cdot\rangle )\Big)({\bf f})\\
				=\sqrt{\frac{1}{\det(\mathbf{Id+L(Id+K)^{-1}})}}
				\exp(-\frac{1}{2} \langle {\bf f}, \mathbf{(Id+K+L)^{-1}} {\bf f} \rangle ),\quad {\bf f} \in S_{d}(\R),
			\end{multline*}	
	in the case the right hand side indeed is a U-funcional.		
\end{example}

\begin{definition}\label{prodnexp}
Let $\mathbf{K}: L^2_{d}(\R, dx)_{\C} \to L^2_{d}(\R, dx)_{\C}$ be as in Definition \ref{D:Nexp}, i.e.~${\rm{Nexp}}(- \frac{1}{2} \langle \cdot ,\mathbf{K} \cdot \rangle)$ exists. Furthermore let $\mathbf{L}: L^2_d(\R, dx)_{\C} \to L^2_d(\R, dx)_{\C}$ be trace class. Then we define 
$$
{\rm{Nexp}}( - \frac{1}{2} \langle \cdot,\mathbf{K} \cdot\rangle ) \cdot \exp( - \frac{1}{2} \langle \cdot,\mathbf{L} \cdot\rangle )$$ via its $T$-transform, whenever 
\begin{multline*}
				T\Big({\rm{Nexp}}( - \frac{1}{2} \langle \cdot,\mathbf{K} \cdot\rangle ) \cdot \exp( - \frac{1}{2} \langle \cdot,\mathbf{L} \cdot\rangle )\Big)({\bf f})\\
				=\sqrt{\frac{1}{\det(\mathbf{Id+L(Id+K)^{-1}})}}
				\exp(-\frac{1}{2} \langle {\bf f}, \mathbf{(Id+K+L)^{-1}} {\bf f} \rangle ),\quad {\bf f} \in S_{d}(\R),
			\end{multline*}	
is a U-functional.
\end{definition}

In the case $\mathbf{g} \in S_d(\R)$, $c\in\C$ the product between the Hida distribution $\Phi$ and the Hida test function $\exp(i \langle \mathbf{g},. \rangle + c)$ can be defined because $(S)$ is a continuous algebra under pointwise multiplication. The next definition is an extension of this product.

\begin{definition}\label{linexp}
The pointwise product of a Hida distribution $\Phi \in (S)'$ with an exponential of a linear term, i.e.~
\begin{equation*}
\Phi \cdot \exp(i \langle {\bf g}, \cdot \rangle  +c), \quad {\bf g} \in L^2_{d}(\R)_{\C}, \, c \in \C,
\end{equation*}
is defined by 
\begin{equation*}
T(\Phi \cdot \exp(i\langle  {\bf g}, \cdot \rangle  + c))({\bf f}):= T\Phi({\bf f}+{\bf g})\exp(c),\quad {\bf f} \in S_d(\R),  
\end{equation*}
if $T\Phi$ has a continuous extension to $L^2_d(\R)_{\C}$ and the term on the right-hand side is a U-functional in ${\bf f} \in S_d(\R)$.
\end{definition}

\begin{definition}\label{donsker}
Let $D \subset \R$ with $0 \in \overline{D}$. Under the assumption that $T\Phi$ has a continuous extension to $L^2_d(\R)_{\C}$, ${\boldsymbol\eta}\in L^2_d(\R)_{\C}$, $y \in \R$, $\lambda \in \gamma_{\alpha}:=\{\exp(-i\alpha)s|\, s \in \R\}$ and that the integrand 
\begin{equation*}
\gamma_{\alpha} \ni \lambda \mapsto \exp(-i\lambda y)T\Phi({\bf f}+\lambda {\boldsymbol\eta}) \in \C
\end{equation*}
fulfills the conditions of Corollary \ref{intcor} for all $\alpha \in D$. Then one can define the product 
\begin{equation*}
\Phi \cdot \delta_0(\langle {\boldsymbol\eta}, \cdot \rangle-y),
\end{equation*}
by
\begin{equation*}
T(\Phi \cdot \delta_0(\langle {\boldsymbol\eta}, \cdot \rangle-y))({\bf f})
:= \lim_{\alpha \to 0} \int_{\gamma_{\alpha}} \exp(-i \lambda y) T\Phi({\bf f}+\lambda {\boldsymbol\eta}) \, d \lambda.
\end{equation*}
Of course under the assumption that the right-hand side converges in the sense of Corollary \ref{seqcor}, see e.g.~\cite{GS98a}.
\end{definition}

This definition is motivated by the definition of Donsker's delta, see Definition \ref{D:Donsker}.

\begin{lemma}{\cite{BG10}}\label{thelemma}
Let  $\mathbf{L}$ be a $d\times d$ block operator matrix on $L^2_{d}(\R)_{\C}$ acting componentwise such that all entries are bounded operators on $L^2(\R)_{\C}$.
Let $\mathbf{K}$ be a d $\times d$ block operator matrix on $L^2_{d}(\R)_{\C}$, such that $\mathbf{Id+K}$ and $\mathbf{N}=\mathbf{Id}+\mathbf{K}+\mathbf{L}$ are bounded with bounded inverse. Furthermore assume that $\det(\mathbf{Id}+\mathbf{L}(\mathbf{Id}+\mathbf{K})^{-1})$ exists and is different from zero (this is e.g.~the case if $\mathbf{L}$ is trace class and -1 in the resolvent set of $\mathbf{L}(\mathbf{Id}+\mathbf{K})^{-1}$).
Let $M_{\mathbf{N}^{-1}}$ be the matrix given by an orthogonal system $({\boldsymbol\eta}_k)_{k=1,\dots J}$ of non--zero functions from $L^2_d(\R)$, $J\in \N$, under the bilinear form $\left( \cdot ,\mathbf{N}^{-1} \cdot \right)$, i.e.~ $(M_{\mathbf{N}^{-1}})_{i,j} = \left( {\boldsymbol\eta}_i ,\mathbf{N}^{-1} {\boldsymbol\eta}_j \right)$, $1\leq i,j \leq J$.
Under the assumption that either 
\begin{eqnarray*}
\Re(M_{\mathbf{N}^{-1}}) >0 \quad \text{ or }\quad \Re(M_{\mathbf{N}^{-1}})=0 \,\text{ and } \,\Im(M_{\mathbf{N}^{-1}}) \neq 0,
\end{eqnarray*} 
where $M_{\mathbf{N}^{-1}}=\Re(M_{\mathbf{N}^{-1}}) + i \Im(M_{\mathbf{N}^{-1}})$ with real matrices $\Re(M_{\mathbf{N}^{-1}})$ and $\Im(M_{\mathbf{N}^{-1}})$, \\
then
\begin{equation*}
\Phi_{\mathbf{K},\mathbf{L}}:={\rm Nexp}\big(-\frac{1}{2} \langle \cdot, \mathbf{K} \cdot \rangle \big) \cdot \exp\big(-\frac{1}{2} \langle \cdot, \mathbf{L} \cdot \rangle \big) \cdot \exp(i \langle \cdot, {\bf g} \rangle)
\cdot \prod_{i=1}^J \delta_0 (\langle \cdot, {\boldsymbol\eta}_k \rangle-y_k),
\end{equation*}
for ${\bf g} \in L^2_{d}(\R,\C),\, t>0,\, y_k \in \R,\, k =1\dots,J$, exists as a Hida distribution. \\
Moreover for ${\bf f} \in S_d(\R)$
\begin{multline}\label{magicformula}
T\Phi_{\mathbf{K},\mathbf{L}}({\bf f})=\frac{1}{\sqrt{(2\pi)^J  \det((M_{\mathbf{N}^{-1}}))}}
\sqrt{\frac{1}{\det(\mathbf{Id}+\mathbf{L}(\mathbf{Id}+\mathbf{K})^{-1})}}\\ 
\times \exp\bigg(-\frac{1}{2} \big(({\bf f}+{\bf g}), \mathbf{N}^{-1} ({\bf f}+{\bf g})\big) \bigg)
\exp\bigg(-\frac{1}{2} (u,(M_{\mathbf{N}^{-1}})^{-1} u)\bigg),
\end{multline}
where
\begin{equation*}
u= \left( \big(iy_1 +({\boldsymbol\eta}_1,\mathbf{N}^{-1}({\bf f}+{\bf g})) \big), \dots, \big(iy_J +({\boldsymbol\eta}_J,\mathbf{N}^{-1}({\bf f}+{\bf g})) \big) \right).
\end{equation*}
\end{lemma}


\section{The Feynman integrand for a Charged Particle in a Constant Magnetic Field} 
In classical mechanics a charged particle moving through a magnetic field $\mathbf{H}=(0,0,H_3)$ has the Lagrangian
$$ L(\bmx, \dot{\bmx}) = \frac{1}{2}m (\dot{\bmx_1}^2 +\dot{\bmx_2}^2  +\dot{\bmx_3}^2) + \frac{q H_3}{c} \left(x_1\dot{x_2}-\dot{x_1}x_2\right),$$
where $m$ is the mass of the particle. We denote the constant in front of the potential term by $k:=\frac{q H_3}{c}$. We see that, beneath the dependence on the spatial coordinates, the potential term depends explicitly on the 
velocities. \\
Since the above three dimensional system can be separated to the free motion parallel to the magnetic field vector and a motion in the plane orthogonal to the magnetic field vector, we restrict ourselves to the two-dimensional system. \\
In the following we realize rigorously the ansatz    
\begin{multline}\label{eqfey}
I_{\rm{mag}} = {\rm Nexp}\left( \frac{i}{\hbar}\int_{0}^t  \frac{\dot{\bmx}(\tau)^2}{2m} d\tau +\frac{1}{2}\int_{t_0}^t \dot{\bmx}(\tau)^2\right) \\ 
\times \exp\left(-\frac{ik}{\hbar} \int_{0}^t \left(x_1(\tau) \dot{x_2}(\tau)-\dot{x_1}(\tau)x_2(\tau)\right) \, d\tau\right) \cdot \delta_0(\bmx(t)-\bmy),
\end{multline}
for the Feynman integrand of a charged particle in a constant magnetic field, with the help of Lemma \eqref{thelemma}. See the introduction for a physical motivation. In \eqref{eqfey} the path $\bmx$ is realized by a two-dimensional Brownian motion starting in $0$ at time $t_0=0$. Then the first term in \eqref{eqfey} can be written as an exponential of quadratic type and gives a generalized Gauss kernel, see Definition \ref{GGK}. Indeed with $\hbar = m =1,$
\begin{equation}\label{Nexpquad}
{\rm Nexp}\left( i \int_0^t  \frac{\dot{\bmx}(\tau)^2}{2} d\tau +\frac{1}{2}\int_0^t \dot{\bmx}(\tau)^2\right) ={\rm Nexp}\left(-\frac{1}{2} \langle (\omega_1 , \omega_2),\mathbf{K} (\omega_1 , \omega_2) \rangle \right),
\end{equation}
with $\mathbf{K} := -(i+1)\mathbf{P}_{[0,t)} := -(i+1) \left(\begin{array}{l l } P_{[0,t)} & 0\\ 0 & P_{[0,t)} \end{array}\right)$, where $P_{[0,t)}$ denotes the orthogonal projection in $L^2(\R)_{\C}$ given by the multiplication with $\1_{[0,t)}$.\\

In the following we derive the desired properties for applying Lemma \ref{thelemma}. First we write also the potential term in \eqref{eqfey} in a quadratic way. 
\begin{proposition}\label{magneticL}
The operator matrix
\begin{align}
\bml=\mathbf{P}_{[0,t)}\left(\begin{matrix} 0 & ik\left(A-A^*\right)\\ik\left(A^* - A\right) & 0\end{matrix}\right)\mathbf{P}_{[0,t)},\label{magnetiLformula}
\end{align} 
fulfills
\begin{align*}
\frac12\langle\mathbf{f},\bml\mathbf{f}\rangle &=-ik\int\limits_{0}^t \left(\int_0^{\tau} f_1(s) \, ds  f_2(\tau)-f_1(\tau) \int_0^{\tau}f_2(s) \, ds \,\right) d\tau, \quad 0\leq t<\infty,
\end{align*}
where $\mathbf{f} = (f_1,f_2) \in L_2^2(\R)$ and operator $A$ is defined by $$A f(\tau) = \1_{[0,t)}(\tau) \int_{[0,\tau)} f(s) \, ds,\quad f \in L^2(\R), \tau\in \R.$$ $A^*$ denotes its adjoint w.r.t.~the bilinear dual pairing $ \langle \cdot,\cdot \rangle$. Moreover $\bml$ is symmetric w.r.t.~$\langle \cdot , \cdot \rangle$.
\end{proposition}

\noindent{\bf Proof:}
With $\bml$ as above we have by the symmetry of the dual pairing
\begin{align*}
\langle\mathbf{f},\bml \mathbf{f}\rangle&=\left\langle\left(\begin{matrix}f_1\\ f_2\end{matrix}\right),
\left(\begin{matrix} 0 & ikP_{[0,t)}\left(A-A^*\right)\\ikP_{[0,t)}\left(A^*-A\right) & 0\end{matrix}
\right)\left(\begin{matrix}f_1\\\ f_2\end{matrix}\right)
\right\rangle\\
&=\left\langle f_1,ikP_{[0,t)}Af_2\right\rangle-\left\langle f_1 , ikP_{[0,t)}A^*f_2\right\rangle\\
&\hspace{35mm}+\left\langle f_2,ikP_{[0,t)}A^*f_1\right\rangle-\left\langle f_2, ikP_{[0,t)}A f_1\right\rangle\\
&=2\left\langle f_1,ikP_{[0,t)} A f_2\right\rangle-2\left\langle f_2,ikP_{[0,t)}A f_1\right\rangle\\
&=2ik\int\limits_{0}^t \left(\int_0^{\tau} f_1(s) \, ds f_2(\tau)-f_1(\tau)  \int_0^{\tau}f_2(s) \, ds\right),
\end{align*}
since $P_{[0,t)}$ and $A$ commute.
\hfill $\blacksquare$

\begin{remark}
If we extend $\langle \cdot, \bml \cdot \rangle$ informally to an element $\boldsymbol{\omega}\in S'_2(\R)$ we have 
\begin{equation}\label{80}
\frac{1}{2} \langle \boldsymbol{\omega}, \bml \boldsymbol{\omega} \rangle = -ik\int\limits_{0}^t B_\tau(\omega_1)\omega_2(\tau)- B_\tau(\omega_2)\omega_1(\tau) \,d\tau,
\end{equation}
where $B_\tau$ is the representation of a one-dimensional Brownian motion as in (4). The term in \eqref{80} corresponds in this case with the potential term in the Feynman integrand \eqref{eqfey}.
\end{remark}

\begin{lemma}\label{magneticinvertability}
The operator $\bmn:L^2_{2}(\R,dx)_{\C}\maps L^2_{2}(\R,dx)_{\C}$ given by
\begin{align*}
\bmn:=\mathbf{P}_{[0,t)}\left(\begin{matrix}
-i Id &ik\left(A-A^*\right)\\
ik\left(A^*-A\right)&-i Id
\end{matrix}\right)\mathbf{P}_{[0,t)}+\mathbf{P}_{[0,t)^c},
\end{align*}
is bijective. Here $A$ and $A^*$ are as in Proposition \ref{magneticeigenbladeterminant}, $0< t< \infty$.
\end{lemma}

\noindent{\bf  Proof:} 
We have
\begin{multline}\label{magneticNdefinition}
\bmn=\bmid+\bmk+\bml\\
\hspace{5 mm}=\mathbf{P}_{[0,t)}\left(\begin{matrix}
-i&ik\left(A-A^*\right)\\
ik\left(A^*-A\right)&-i\\
\end{matrix}\right)\mathbf{P}_{[0,t)}+\mathbf{P}_{[0,t)^c}
=-i\mathbf{P}_{[0,t)}\bmn_1 \mathbf{P}_{[0,t)}+\mathbf{P}_{[0,t)^c}.\\
\end{multline}

\noindent Denote the restriction of $\mathbf{P}_{[0,t)} \bmn_1$ to $L^2_2([0,t),\R)_{\C}$ by $\bmn_2$. Then bijectivity of $\bmn_2$ implies bijectivity of $\bmn$ and 
$$ \bmn^{-1} = i \bmn_2^{-1}\mathbf{P}_{[0,t)} + \mathbf{P}_{[0,t)^c}.$$ 
For this we show that $\bmn_2$ is Fredholm with $\ker(\bmn_2)=\{0\}$. 

First we show that 
$$\left(\begin{array}{l l} 0 & -k (A-A^* ) \\ -k( A^* -A ) & 0 \end{array}\right)=:\left(\begin{array}{l l} 0 & M \\ M^* & 0 \end{array}\right) $$ is compact on $L^2_2 ([0,t),\R)_{\C}$. Then we have 
$$\bmn_2 = \left(\begin{array}{l l} Id & M \\ M^* & Id \end{array} \right),$$
is a compact perturbation of the identity on $L^2_2 ([0,t),\R)_{\C}$. We have

\begin{multline*}
\left(Af\right)(\tau)=\int\limits_{0}^{\tau} f(s)\,ds=\int\limits_{0}^t\1_{[0,\tau)}(s)f(s)\,ds\, \\
\text{and }\,\left(A^*f\right)(\tau)=\int\limits_\tau^t f(s)\,ds=\int\limits_{0}^t\1_{[\tau,t)}(s)f(s)\,ds.
\end{multline*}

If $\1_{[0,\tau)}$ and $\1_{[\tau,t)}$ are Hilbert-Schmidt-kernels, the above integral operators $A$ and $A^*$ are compact operators on $L^2([0,t),dx)_{\C}$ and so are $M$ and $M^*$. Indeed

\begin{align*}
&\int\limits_{0}^t\int\limits_{0}^t (\1_{[0,s)}(\tau))^2 \, d\tau ds=\int\limits_{0}^t\int\limits_{0}^t (\1_{[s,t)}(\tau))^2 \, d\tau ds=\frac12 t^2<\infty.
\end{align*}
$M$ as well as $M^*$can be written as the limit of a sequence of finite rank operators $(M_n)_{n\in \N}$ and $(M_n^*)_{n\in \N}$, respectively, in operator norm. Then:
\begin{multline*}
\sup_{\|(f_1 , f_2)\|\leq 1} \left\| \Bigg(\left(\begin{array}{l l} 0 & M \\ M^* & 0 \end{array}\right)- \!\left(\begin{array}{l l} 0 & M_n \\ M_n^* & 0 \end{array}\right)\Bigg)\left(\begin{array}{l} f_1 \\ f_2 \end{array} \right)\right\|\\
\leq\!\sup\limits_{\|f_1\| \leq1}\|(M-M_n)f_1\|+\!\sup\limits_{\|f_2\| \leq1}\!\|(M^*-M^*_n)f_2\|,
\end{multline*} 
where the right hand side tends to zero as $n$ goes to $\infty$. Hence, $\left(\begin{array}{l l} 0 & M \\ M^* & 0 \end{array}\right)$ as the limit of finite rank operators is compact.\\
It is left to show that $\ker\left(\bmn_2\right)=\{0\}$. Let

\begin{align*}
\left(\begin{matrix}
Id&M\\
M^*&Id
\end{matrix}\right)
\left(\begin{matrix}
f_1\vphantom{\left(A-A^*\right)}\\f_2\vphantom{\left(A-A^*\right)}
      \end{matrix}
\right)=
\left(\begin{matrix}
0\vphantom{\left(A-A^*\right)}\\0\vphantom{\left(A-A^*\right)}
\end{matrix}
\right).
\end{align*}

This leads to the system
\begin{align*}
f_1(s)&=k\int\limits_{0}^sf_2(\tau)d\tau-k\int\limits_s^tf_2(\tau)d\tau,\quad s \in [0,t),\\
f_2(s)&=k\int\limits_s^tf_1(\tau)d\tau-k\int\limits_{0}^sf_1(\tau)d\tau,\quad s \in [0,t).
\end{align*}

An analogue calculation as in the proof of Proposition \ref{magneticeigenbladeterminant}, below, yields $f_1\equiv f_2\equiv0$, which gives $\ker\left(\bmn_2\right)=\{0\}$.
\hfill $\blacksquare$

Now we want to determine the prefactor in Equation \eqref{magicformula}. Recall that the determinant of a diagonalizable operator is defined as the product of its eigenvalues, if it exists. We have the following proposition.

\begin{proposition}\label{magneticeigenbladeterminant}
Let $\bmk$ be as in \eqref{Nexpquad}, $\bml$ as in Proposition \ref{magneticL}. Then
\begin{itemize}
\item[(i)] For $\bml(\bmid+\bmk)^{-1}:L^2_{2}(\R,dx)_{\C}\maps L^2_{2}(\R,dx)_{\C}$, the non-vanishing eigenvalues and their corresponding eigenvectors are
\begin{align*}
\lambda_n&=\frac{2k}{(2n-1)\pi}t,\\
e_n(\cdot)&=c_1\left(\begin{matrix}
\1_{[0,t)}(\cdot)\cos\left(\frac{2k}{\lambda_n}\cdot\right)\\
\1_{[0,t)}(\cdot)\sin\left(\frac{2k}{\lambda_n}\cdot\right)
\end{matrix}\right)+c_2\left(\begin{matrix}
\1_{[0,t)}(\cdot)\sin\left(\frac{2k}{\lambda_n}\cdot\right)\\
-\1_{[0,t)}(\cdot)\cos\left(\frac{2k}{\lambda_n}\cdot\right)
\end{matrix}\right),
\end{align*}
$n\in\Z,c_1,c_2\in\C$, where the multiplicity of the eigenvalues is 2.
\item[(ii)] We have for the determinant
\begin{align*}
\det\left(\bmid+\bml(\bmid+\bmk)^{-1}\right)=\cos^2\!\left(kt\right).
\end{align*}
\end{itemize}
\end{proposition}

\noindent{\bf Proof:}
(i): We want to calculate the eigenvalues of
\begin{align*}
&\bml(\bmid+\bmk)^{-1}\\[3mm]
=&\left(\begin{matrix}
0 & ikP_{[0,t)}\left(A-A^*\right)P_{[0,t)}\\
ikP_{[0,t)}\left(A^*-A\right)P_{[0,t)} & 0\end{matrix}
\right)\\
&\hspace{55mm}\times
\left(\begin{matrix}
iP_{[0,t)}+P_{[0,t)^c}&0\vphantom{ikP_{[0,t)}\left(A-A^*\right)}\\
0\vphantom{ikP_{[0,t)}\left(A-A^*\right)}&iP_{[0,t)}+P_{[0,t)^c}
             \end{matrix}\right)\\[5mm]
=&\left(\begin{matrix}
0&kP_{[0,t)}\left(A^*-A\right)P_{[0,t)}\\
kP_{[0,t)}\left(A-A^*\right)P_{[0,t)}&0
        \end{matrix}\right) = \left( \begin{matrix}
0&M\\
M^*&0
   \end{matrix}\right)\mathbf{P}_{[0,t)},
\end{align*}
with $M$ and $M^*$, as in Lemma \ref{magneticinvertability}, respectively.  
Hence we interpret the operator matrix as an operator from $L^2_{2}([0,t),dx)_\C$ into itself and restrict the desired eigenfunctions from now on to this interval. Assume
\begin{align*}
\left(\begin{matrix}
0&M\\
M^*&0
        \end{matrix}\right)
\left(\begin{matrix}
e_{n,1}\vphantom{kP_{[0,t)}\left(A-A^*\right)}\\
e_{n,2}\vphantom{kP_{[0,t)}\left(A-A^*\right)}
      \end{matrix}\right)=\lambda_n\left(\begin{matrix}
e_{n,1}\vphantom{kP_{[0,t)}\left(A-A^*\right)}\\
e_{n,2}\vphantom{kP_{[0,t)}\left(A-A^*\right)}
      \end{matrix}\right), \quad \left(\begin{matrix} e_{n,1} \\e_{n,2} \end{matrix} \right)\in L^2_2([0,t), dx)_{\C}, n \in \Z.
\end{align*}

Then
\begin{align}
&k\left(A^*-A\right)e_{n,2}=\lambda_ne_{n,1}(\cdot)\label{41}\\
\text{and}\hspace{3mm}&k\left(A-A^*\right)e_{n,1}=\lambda_ne_{n,2}(\cdot)\label{42}.
\end{align}

Differentiation yields
\begin{equation*}
-2ke_{n,2}=\lambda_ne'_{n,1} \quad \text{ and }\quad 2ke_{n,1}=\lambda_ne'_{n,2}.
\end{equation*}

Hence

\begin{align*}
e''_{n,1}+\frac{4k^2}{\lambda_n^2}e_{n,1}=0 \quad \text{ and }\quad e''_{n,2}+\frac{4k^2}{\lambda_n^2}e_{n,2}=0.
\end{align*}

Due to general theory on ordinary differential equations, the solutions read

\begin{align}
e_{n,1}(\cdot)&=c_1\cos\left(\frac{2k}{\lambda_n}\cdot\right)+c_2\sin\left(\frac{2k}{\lambda_n}\cdot\right)\label{43}\\
\text{and}\ \ e_{n,2}(\cdot)&=d_1\cos\left(\frac{2k}{\lambda_n}\cdot\right)+d_2\sin\left(\frac{2k}{\lambda_n}\cdot\right)\label{44},
\end{align}
$c_1,c_2,d_1,d_2\in\C$. Inserting this into the integral Equation (\ref{42}) we get

\begin{align*}
e_{n,2}(s)&=\frac k{\lambda_{n}}\int\limits_{0}^sc_1\cos\left(\frac{2k}{\lambda_n}\tau\right)+c_2\sin\left(\frac{2k}{\lambda_n}\tau\right)d\tau\\*
&\hspace{10mm}-\frac k{\lambda_n}\int\limits_{s}^tc_1\cos\left(\frac{2k}{\lambda_n}\tau\right)+c_2\sin\left(\frac{2k}{\lambda_n}\tau\right)d\tau,\quad s \in [0,t),
\end{align*}
which leads to
\begin{align*}
e_{n,2}(s)&=c_1\sin\left(\frac{2k}{\lambda_n}s\right)-c_2\cos\left(\frac{2k}{\lambda_n}s\right)\\*
&\hspace{4mm}+c_2\left(\cos\left(\frac{2k}{\lambda_n}t\right)+1\right)-c_1\sin\left(\frac{2k}{\lambda_n}t\right),\quad s \in [0,t).
\end{align*} 

Since $e_{n,2}$ is of the form (\ref{44}) we have $d_1=-c_2$, $d_2=c_1$ and

\begin{align}
c_2\left(\cos\left(\frac{2k}{\lambda_n}t\right)+1\right)-c_1\sin\left(\frac{2k}{\lambda_n}t\right)=0\label{45}.
\end{align}
\\[2mm]
Now of course (\ref{41}) must also hold for $s=0$, thus

$$
\lambda_nc_1=\lambda_ne_{n,1}(0)=k\int\limits_{0}^tc_1\sin\left(\frac{2k}{\lambda_n}\tau\right)-c_2\cos\left(\frac{2k}{\lambda_n}\tau\right)d\tau,
$$
which implies
\begin{align}
c_1&\left(\cos\left(\frac{2k}{\lambda_n}t\right)+1\right)=-c_2\sin\left(\frac{2k}{\lambda_n}t\right)\label{46}.
\end{align}

\noindent First assume $c_1=0$, then we have with \eqref{45}
$$c_2\left(\cos\left(\frac{2k}{\lambda_n}t\right)+1\right)=0,$$ 
and with \eqref{46} 
$$-c_2\sin\left(\frac{2k}{\lambda_n}t\right)=0.$$
But as we assume $(e_{n,1},e_{n,2})^T$ to be an eigenvector, the functions $e_{n,1}$ and $e_{n,2}$ may not both be the zero function, i.e.~$c_2\neq 0$.\\
Hence we have
$$\sin\left(\frac{2k}{\lambda_n}t\right)=0 \text{ and }\cos\left(\frac{2k}{\lambda_n}t\right) = -1,$$
which is equivalent to 
\begin{align*}
\frac{2k}{\lambda_n}t=(2n-1)\pi,
\end{align*}
for some $n\in\mathbb{Z}$, i.e.~
\begin{align*}
\lambda_n=\frac{2k}{(2n-1)\pi}t,\ n\in\Z.
\end{align*}

\noindent If we assume $c_2=0$, then 

\begin{equation*}
c_1\sin\left(\frac{2k}{\lambda_n}t\right)=0=c_1\left(\cos\left(\frac{2k}{\lambda_n}t\right)+1\right).
\end{equation*}
This again is equivalent to
\begin{align*}
\frac{2k}{\lambda_n}t=(2n-1)\pi,
\end{align*}
for some $n\in\mathbb{Z}$, i.e.~,
\begin{align*}
\lambda_n=\frac{2k}{(2n-1)\pi}t,\ n\in\Z.
\end{align*}

\noindent Assume $c_1\neq0\neq c_2$, then 
we multiply \eqref{45} on both sides with \eqref{46} and obtain:
$$ c_1  c_2 \left(\cos^2\left(\frac{2k}{\lambda_n}t\right) + 2 \cos\left(\frac{2k}{\lambda_n}t\right) +1 \right) = - c_1 c_2 \sin^2\left(\frac{2k}{\lambda_n}t\right)$$ 
which gives 
$$2 \cos\left(\frac{2k}{\lambda_n}t\right) = - \left(\sin^2\left(\frac{2k}{\lambda_n}t\right) +\cos^2\left(\frac{2k}{\lambda_n}t\right)\right)-1.$$
Thus again 
$$\cos\left(\frac{2k}{\lambda_n}t\right) = -1.$$
Inserted in \eqref{45}, we also obtain
$$
\sin\left(\frac{2k}{\lambda_n}t\right)=0.
$$

At first sight (\ref{45}) and (\ref{46}) give restrictions to the choice of $c_1$ and $c_2$. But naturally, if we have an eigenvector consisting of the two functions $e_{n,1}$ and $e_{n,2}$, corresponding to a certain $\lambda_n$, the factors of $c_1$ and $c_2$ in (\ref{45}) and (\ref{46}) become zero and the aforementioned can be choosen arbitrary. So an eigenfunction to the eigenvalue $\lambda_n$ is always of the form

\begin{align*}
s \mapsto e_n(s)=\left(\begin{matrix}
e_{n,1}(s)\\
e_{n,2}(s)
             \end{matrix}\right)&
=c_1\left(\begin{matrix}
\cos\left(\frac{2k}{\lambda_n}(s)\right)\\
\sin\left(\frac{2k}{\lambda_n}(s)\right)
\end{matrix}\right)+c_2\left(\begin{matrix}
\sin\left(\frac{2k}{\lambda_n}(s)\right)\\
-\cos\left(\frac{2k}{\lambda_n}(s)\right)
\end{matrix}\right),\\
\end{align*}

\noindent where $c_1,c_2\in\C$ are arbitrary and the involved vectors are clearly linearly independent. Thus the dimension of the eigenspace corresponding to $\lambda_n$ and therewith its multiplicity is 2.\\[0.2cm]

(ii): In (i) we calculated the eigenvalues and eigenfunctions of $\mathbf{L}(\bmid +\bmk)^{-1}$ considered as an operator from $L^2_2([0,t),dx)_{\C}$ to itself. The eigenfunctions form a basis of $L^2_2([0,t),dx)_{\C}$, but surely not of $L^2_2(\R,dx)_{\C}$. However, we can extend the set of eigenfunctions to a basis of $L^2_2(\R,dx)_{\C}$ by adding an arbitrary basis $L^2_2([0,t)^c,dx)_{\C}$. Note that because of the projection on $[0,t)$ in $\mathbf{L}(\bmid +\bmk)^{-1}$ all basis functions of $L^2_2([0,t)^c,dx)_{\C}$ are eigenvectors to the eigenvalue $0$. Since the spectrum of $\bmid +\mathbf{L}(\bmid +\bmk)^{-1}$ is just shifted by $1$. This part does not give a contibution to the determinant (multiplication by $1$).\\
Note for the nonvanishing eigenvalues of $\mathbf{L}(\bmid +\bmk)^{-1}$ we have 
\begin{equation*}
\lambda_n=-\lambda_{-n+1}, \quad \text{for all } n\in\Z,
\end{equation*}

\noindent thus
\begin{align*}
(1+\lambda_n)(1+\lambda_{-n+1})=1-\lambda_n^2.
\end{align*}\\
Finally

\begin{multline*}
\det\left(\bmid+\bml(\bmid+\bmk)^{-1}\right)=\prod\limits_{n\in\Z}(1+\lambda_n)^2\\
= \prod\limits_{n\in\N} (1+\lambda_n)(1+\lambda_{-n+1})=\prod\limits_{n\in\N}\left(1-\frac{4k^2}{(2n-1)^2\pi^2}t^2\right)^2=\cos^2\!\left(kt\right).
\end{multline*}
\hfill $\blacksquare$\\

In the following we calculate the preimages of ${\boldsymbol \eta}_1=\left(\begin{matrix}
\1_{[0,t)}\\
0
\end{matrix}\right)$ and ${\boldsymbol \eta}_2=\left(\begin{matrix}
0\\
\1_{[0,t)}
\end{matrix}\right)$ under $N$. With the help of this we can obtain $\left({\boldsymbol \eta_i}, \mathbf{N^{-1}} {\boldsymbol \eta_j}\right)$, $1\leq i,j\leq2$, and hence the matrix $M_{\mathbf{N^{-1}}}$ used in Equation \eqref{magicformula}. 

\begin{proposition}\label{P:Preimage}
Let $\bmn$ as in equation (\ref{magneticNdefinition}). Then
\begin{align*}
\text{(i)}\quad\bmni\left(\begin{matrix}
\1_{[0,t)}\\
0
\end{matrix}\right)&=\left(\begin{matrix}
f_1\\
f_2\end{matrix}\right)=\bmf\in L^2_{2}([0,t),dx)_{\C},\\
\text{(ii)}\quad\bmni\left(\begin{matrix}
0\\
\1_{[0,t)}
\end{matrix}\right)&=\left(\begin{matrix}
g_1\\
g_2\end{matrix}\right)=\bmg\in L^2_{2}([0,t),dx)_{\C},
\end{align*}
with
\begin{align}
f_1(s)&:=i\cos\left(2ks\right)+i\frac{\sin\left(2kt\right)}{\cos\left(2kt\right)+1}\sin\left(2ks\right)=:g_2(s)\label{f1},\quad s \in [0,t)\\
f_2(s)&:=i\frac{\sin\left(2kt\right)}{\cos\left(2kt\right)+1}\cos\left(2ks\right)-i\sin\left(2ks\right)=-:g_1(s)\label{f2},\quad s \in [0,t).
\end{align}
\end{proposition}

\noindent{\bf Proof:}
We have to check that  
\begin{align*}
-i\left(\begin{matrix}
Id&M\\
M^*&Id
\end{matrix}\right)
\left(\begin{matrix}
f_1\vphantom{\left(A-A^*\right)}\\f_2\vphantom{\left(A-A^*\right)}
      \end{matrix}
\right)=
\left(\begin{matrix}
\1_{[0.t)}\vphantom{\left(A-A^*\right)}\\0\vphantom{\left(A-A^*\right)}
\end{matrix}
\right),
\end{align*}
see the proof of Lemma \ref{magneticinvertability}. The corresponding system of equations reads

\begin{align}
-if_1+ik\left(Af_2 -A^*f_2\right)&=1\label{48}\\
ik\left(A^*f_1-Af_1\right)-if_2&=0\label{49}.
\end{align}

Let $s \in [0,t)$, then

$$\left((A- A^*)\sin(2k\cdot)\right)(s) = \int_0^s \sin(2k\tau) \, d\tau -\int_s^t \sin(2k\tau) \, d\tau  =-\frac{\cos(2ks)}{k} + \frac{1+ \cos(2kt)}{2k}, $$
and
$$\left((A- A^*)\cos(2ks)\right)(s) = \int_0^s \cos(2k\tau) \, d\tau - \int_s^t \cos(2k\tau) \, d\tau =  \frac{\sin(2ks)}{k} - \frac{\sin(2kt)}{2k}. $$

Thus
\begin{multline*}
\left((A- A^*) f_2\right)(s) \\
= i\frac{\sin\left(2kt\right)}{\cos\left(2kt\right)+1} \left((A- A^*)\cos\left(2k\cdot\right)\right)(s) -i\left((A- A^*)\sin\left(2k\cdot\right)\right)(s) \\
									= i(\frac{\sin\left(2kt\right)}{\cos\left(2kt\right)+1} (\frac{\sin(2ks)}{k} - \frac{\sin(2kt)}{2k}) +\frac{\cos(2ks)}{k} - \frac{1+ \cos(2kt)}{2k}).\\
\end{multline*}
So we get
\begin{multline*}
-if_1(s)+ik\left(Af_2(s) -A^*f_2(s)\right)\\
= \cos(2ks) + \frac{\sin\left(2kt\right)}{\cos\left(2kt\right)+1}\sin\left(2ks\right) -\frac{\sin\left(2kt\right)}{\cos\left(2kt\right)+1}\sin(2ks) \\
+ \frac{\sin^2(2kt)}{2\cos\left(2kt\right)+1} - \cos(2ks) + \frac{1+ \cos(2kt)}{2} = 1.$$
\end{multline*}
Furthermore
\begin{multline*}
\left((A^*-A) f_1\right)(s)\\
=i\left((A^*- A)\cos\left(2k\cdot\right)\right)(s)+i\frac{\sin\left(2kt\right)}{\cos\left(2kt\right)+1}\left((A^*- A)\sin\left(2k\cdot\right)\right)(s)\\
= \frac{i}{k}\left( \frac{\sin(2kt)}{\cos(2kt)+1} \cos(2ks) -\sin(2ks) \right).
\end{multline*}
And hence we obtain
\begin{multline*}
-ik\left(Af_1(s)-A^*f_1(s)\right)-if_2(s) \\
= -\frac{\sin(2kt)}{\cos(2kt)+1} \cos(2ks) +\sin(2ks) +\frac{\sin(2kt)}{\cos(2kt)+1} \cos(2ks) -\sin(2ks)=0.
\end{multline*}
Thus (i) is shown. (ii) can be shown analogously. 
An analogue computation also yields for $\mathbf{g}$.

\hfill $\blacksquare$

Now all conditions of Lemma \ref{thelemma} are fulfilled. Hence we have the following theorem.

\begin{theorem}[Feynman integrand for a charged particle in a magnetic field]\label{magnetictheorem}
Let $0\leq 0< t<\infty$ with
\begin{align*}
\frac{kt}{\pi},\frac{kt}{\pi}+\frac{1}{2} \notin \Z.
\end{align*}
Then the Feynman integrand $I_{\rm{mag}}$ for a charged particle in a constant the magnetic field exists as a Hida Distribution. Moreover the integrand can be written as
\begin{equation*}
I_{\rm{mag}}=\Nexp\left(-\frac{1}{2} \langle {\boldsymbol \omega},\mathbf{K} {\boldsymbol \omega} \rangle\right)\cdot\exp\left(-\frac{1}{2} \langle {\boldsymbol \omega},\bml {\boldsymbol \omega} \rangle\right)\cdot\delta_0\left(\mathbf{B}_{t} -{\bmy}\right),
\end{equation*}
where $\bmy=(y_1,y_2)^T\in\R^2$ and the operators $\mathbf{K}$ as in \eqref{Nexpquad} and $\bml$ as in Proposition \ref{magneticL}.Its $T$-transform in $\boldsymbol{\varphi} \in S_2(\R)$ is given by
\begin{multline}\label{magneticTtransform}
TI_{mag}(\boldsymbol{\varphi})\notag=\frac{k}{2\pi i}\frac{1}{\cos(\left(kt\right)}\exp\left(-\frac12\left(\boldsymbol{\varphi},\bmni\boldsymbol{\varphi}\right)\right)\\
\times\exp\left(-\frac{ik}{2}\cot\left(kt\right)\Bigg(\left(iy_1+\frac12\left(\1_{[0,t)},(\bmni\boldsymbol{\varphi})_1\right)+\frac12\left(\boldsymbol{\varphi},\mathbf{f}\right)\right)^2\right.\\
\hspace{45mm}\left.\left.+\left(iy_2+\frac12\left(\1_{[0,t)},(\bmni\boldsymbol{\varphi})_2\right)+\frac12\left(\boldsymbol{\varphi},\mathbf{g}\right)\right)^2\Bigg)\right)\right.,
\end{multline}
for all $\boldsymbol{\varphi}\in S_2(\R)$. Here $\bmf=(\1_{[0,t)} f_1,\1_{[0,t)} f_2)^T,\bmg=(\1_{[0,t)} g_1,\1_{[0,t)} g_2)^T$ with
\begin{align*}
f_1(s)&=i\cos\left(2kt\right)+i\frac{\sin\left(2kt\right)}{\cos\left(2kt\right)+1}\sin\left(2ks\right)=g_2(s), \quad s \in [0,t)\\
f_2(s)&=i\frac{\sin\left(2kt\right)}{\cos\left(2kt\right)+1}\cos\left(2ks\right)-i\sin\left(2ks\right)=-g_1(s),\quad s \in [0,t).
\end{align*}
The generalized expectation ($T$-transform in $\boldsymbol{\varphi}=0$) gives 
\begin{equation}\label{magneticpropagator}
TI_{mag}(0)
=\frac{k}{2\pi i}\frac{1}{\cos\left(kt\right)}\exp\left(\frac{ik}{2}\cot\left(kt\right)\left(y_1^2+y_2^2 \right)\right),
\end{equation}
which coincides with the Greens function for a charged particle in a magnetic field see e.g.~\cite{KL85}, \cite{G96}.
\end{theorem}

\noindent {\bf Proof:}
By Proposition \ref{P:Preimage} we have that $M_{\bmn^{-1}}$ is completely imaginary and thus fulfills the conditions of Lemma \ref{thelemma}. The prefactor in the exponential function in \eqref{magicformula} exists whenever the $\cot(kt) \neq \infty$, which is for $kt \neq n \pi , n \in \Z.$
Furthermore $\bmn$ is invertible by Theorem \ref{magneticinvertability}. By Proposition \ref{magneticeigenbladeterminant} we have the the determinant of $\bmid +\bml(\bmid +\bmk)^{-1}$ exists and the prefactor in \eqref{magicformula} is finite whenever the $\cos{kt} \neq 0$, i.e.~$kt\neq n + \frac{1}{2}$, for $n \in \N$. 
Hence we have that the conditions of Lemma \ref{thelemma} are fulfilled and 
\begin{equation*}
I_{\rm{mag}}=\Nexp\left(-\frac{1}{2} \langle {\boldsymbol \omega},\mathbf{K} {\boldsymbol \omega} \rangle\right)\cdot\exp\left(-\frac{1}{2} \langle {\boldsymbol \omega},\bml {\boldsymbol \omega} \rangle\right)\cdot \delta_0\left(\mathbf{B}_{t} -{\bmy}\right),
\end{equation*}
is a Hida disribution. The $T$-transform is provided by Lemma \ref{thelemma}
\hfill $\blacksquare$

\begin{remark}\label{magneticjusitfication}
At the critical time $t$, with $\frac{kt}{\pi}\in\Z$ or $\frac{kt}{\pi}\in\Z+\frac{1}{2}\Z$  the Feynman propagator again is the Dirac delta function at 0. In the theory of Maslov (Morse) correction this singularity is called
caustics, see e.g.~\cite{GS98a} Remark 5.2 and \cite{S81}. Another typical example for caustics beneath a charged particle in a constant magnetic field is the harmonic oscillator, \cite{GS98a}. Note that the Greens function for small times always exists.
\end{remark}

\begin{remark}
In this article we considered the charged particle in a magnetic field without any electric induced force. The system with an external electric force $\mathbf{F} \in S_2^{\infty}(\R)$ is represented by the Lagrangian
$$
L(\bmx, \dot{\bmx}) = \frac{1}{2}m (\dot{\bmx_1}^2 +\dot{\bmx_2}^2  +\dot{\bmx_3}^2) + \frac{q H_3}{c} \left(x_1\dot{x_2}-\dot{x_1}x_2\right) - \mathbf{\dot{F}}\mathbf{x}.
$$
The Greens function for this system can be obtained by considering $T(I_{mag})(\mathbf{F})$, see e.g.~\cite{HKPS93,G96,GV08}.
\end{remark}

\end{document}